# Using artificial neural networks to improve photometric modeling in airless bodies

J. L. Rizos[a,b], A. Asensio Ramos[a,b], R. Golish[c], D. N. DellaGiustina[c], J. Licandro[a,b], J. de León[a,b], H. Campins[d], E. Tatsumi[a,b], and M. Popescu[e]

a Instituto de Astrofísica de Canarias, C/Vía Láctea s/n, E-38205 La Laguna, Tenerife, Spain

b Departamento de Astrofísica, Universidad de La Laguna, E-38206 La Laguna, Tenerife, Spain

c Lunar and Planetary Laboratory, University of Arizona, 1415 N. Sixth Ave., Tucson, AZ 85705-0500, USA

d Department of Physics, University of Central Florida, P.O. Box 162385, Orlando, FL 32816-2385, USA

e Astronomical Institute of the Romanian Academy, 5 Cuţitul de Argint, 040557 Bucharest, Romania

## Highlights

- We present an artificial neural network for photometric modeling.

- We validate our method using (101955) Bennu images of the OSIRS-REx mission acquired by MapCam.

- Our approach provides a more precise modeling for all color filters, offering an improvement of up to 14.30%, as well as a considerable reduction in time.

## Abstract

Relevant information about physical properties of the surface of airless bodies such as porosity, particle size, or roughness can be inferred knowing the dependence of the brightness with illumination and observing geometry. Additionally, this knowledge is necessary to standardize or photometrically correct data acquired under different illumination conditions. In this work we develop a robust, automatic, and efficient photometric modeling methodology, which does not start from preliminary assumptions that may bias the analysis, and we tested and validated it using Bennu images acquired by the camera MapCam from the OSIRIS-REx spacecraft. It consists of a supervised machine learning algorithm through an artificial neural network. Our system provides a more precise modeling for all color filters than the previous procedures which are already published, offering an improvement over this classic approach of up to 14.30%, as well as a considerable reduction in computing time.

*Keywords: Photometric modeling; Machine Learning; (101955) Bennu; OSIRIS-REx*

# 1. Introduction

Studying the dependence of the brightness with illumination and observing geometry in airless bodies is useful for multiple reasons. First, it allows us to retrieve relevant information about physical properties of the surface. There are many characteristics that will condition the reflected light on the surface such as porosity, particle size, or roughness, which can be inferred by analyzing the reflected light. One needs to standardize observing geometries to compare data acquired under different illumination conditions. This standardization procedure, commonly called photometric correction, consists of finding a mathematical model that describes how light is reflected as a function of the geometric phase ($\alpha$), emission ($e$), and incidence ($i$) angles. For an arbitrary point on a surface **P**, illuminated from a point **S** and observed from another point **C** (Figure 1), the phase angle is defined as the intersection angle between the two lines **PS** and **PC**, the emission angle as that formed by the segment **PC** and the normal to that point **n**, and the angle of incidence as the angle between **PS** and **n**. Then, by means of a model, it is possible to translate the measured brightness under any geometric values ($\alpha_m, e_m, i_m$) to a reference ones ($\alpha_0, e_0, i_0$) (Eq. 1). The most common reference angles used in photometric corrections are ($\alpha_0, e_0, i_0$) = (0°, 0°, 0°), which correspond to the normal reflectance, or ($\alpha_0, e_0, i_0$) = (30°, 0°, 30°), which is a common laboratory setting (also known as standard reflectance). Hereinafter we refer to them as normal or laboratory reference, respectively.

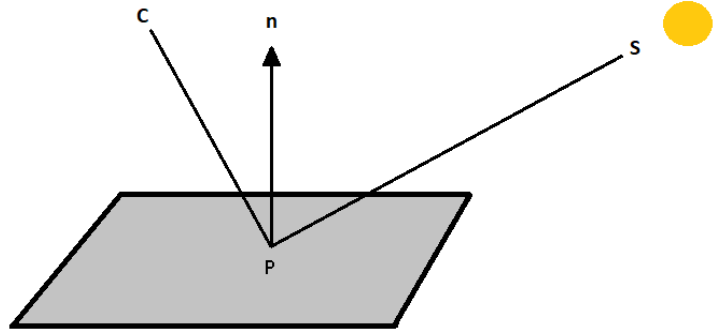

*Figure 1. Diagram of the positions of the camera C, the Sun S and a point P on a surface. The reflectance value will depend on the relative positions of these three elements.*

$$RADF(\alpha_0, e_0, i_0) = RADF(\alpha_m, e_m, i_m) \frac{RADF_{model}(\alpha_0, e_0, i_0)}{RADF_{model}(\alpha_m, e_m, i_m)} \quad (1)$$

There are several mathematical models (also called photometric models) that describe reflectance as a function of the phase, incidence, and emission angles. These include Hapke (Hapke 1963, 1981, 1986, 1993, 2012) or Shkuratov models (Shkuratov et al. 1999), which are physically motivated and based on the radiative transfer theory. Moreover, there are also purely empirical models such as Lambert or Minnaert, that have been developed *ad hoc* (Li et al. 2015). The theoretical origin of Hapke model allows for a physical interpretation of its terms and thus it is a popular approach. However, there are several cases where correlations between supposedly independent Hapke parameters have been found (Gunderson et al. 2006; Shepard & Helfenstein 2007; Shkuratov et al. 2012). Likewise, empirical models are based on *a priori* dependence relationships between the geometric angles, which can bias photometric analysis because they might just be approximating more complex relationships. Empirical approaches assume that the photometric model can be decomposed as the product of two functions. A phase function, dependent only on the phase, and a disk function, which will depend on the emission and incidence values, and for some models also on the phase. Limitations to the existing physically motivated and empirical techniques have prompted us to seek new approaches. Our objective in this work is to find new way that could either replace or complement the classic approach to photometric modelling.

Currently, planetary science community is witnessing unprecedent growth in robotic exploration, a field where photometric modeling is an indispensable tool. Recent missions to airless bodies such as OSIRIS-REx (Lauretta et al. 2017) or Hayabusa2 (Watanabe et al. 2017), or those planned in the short term such as Martian Moons eXploration – MMX (Kuramoto et al. 2018), would benefit considerably from having rapid tools for establishing photometric models in real time. The exponential growth of technology, and the possibility of acquiring large-scale data sets, makes manual processing carried out by humans exceedingly challenging.

In this work we develop a robust, automatic, and efficient photometric modeling methodology, making use of a supervised machine learning algorithm. Specifically, we develop an artificial neural network given their powerful application to analogue problems within astronomy (Cambioni et al. 2019; Asensio-Ramos et al. 2021; Moseley et al. 2020). At the same time, we will reproduce classical modeling to compare results. We use OSIRIS-REx data of near-Earth asteroid (101955) Bennu, one of the more recent missions to an airless body of our Solar System. The spacecraft was launched in September 2016 and surface-resolved images have been acquired since October 2018. OSIRIS-REx images are public and available at the Planetary Data System[1]. For this work, we focus on multispectral data collected by the MapCam medium-field-of-view imager. MapCam has a 125-mm focal length and a focal ratio of f/3.3, which provided a ~ 4° field of view (Rizk et al., 2018). It has four narrow band filters based on the Eight-Color Asteroid Survey (ECAS, Tedesco et al., 1982): *b'*, *v*, *w*, and *x*, with effective wavelengths at 473, 550, 698, and 847 nm, respectively. This instrument has taken thousands of images from multiple viewing geometries of asteroid Bennu for each color filter. In Section 2 we present both the classical approach and the new method using machine-learning, as well as our criterion to select and train data and evaluate results. In Section 3 we present and discuss the results of the modelling efficiency. In Section 4 we summarize our conclusions and the planned future work.

---

[1] https://pds.nasa.gov/

## 2. Methods

### 2.1 Classical approach

The classical approach followed so far for photometric modeling consisted of fitting empirical data to a photometric model with several adjustable parameters. This process determines the value of these parameters that minimizes the error. As a reference for this first part, we base our procedure in the conclusions of the work carried out by Golish et al. (2021a), in which, after testing several photometric models, they concluded that the ROLO phase function, $A(\alpha)$, — which includes an explicit term to model the opposition surge— with the Lommel-Seeliger disk function, $d(e, i)$, is the best option to describe how light is reflected from the surface of Bennu (Eqs. (2) – (4)).

$$RADF_{model} = A(\alpha) \cdot d(e, i) \qquad (2)$$

$$A(\alpha) = C_0 e^{-C_1 \alpha} + A_0 + A_1 \alpha + A_2 \alpha^2 + A_3 \alpha^3 + A_4 \alpha^4 \qquad (3)$$

$$d(e, i) = \frac{\cos(i)}{\cos(i) + \cos(e)} \qquad (4)$$

To do this, as in the cited work, we use the color filter images of the Approach, Preliminary survey, and Equatorial stations phases, from October 29th, 2018 to June 6th, 2019. It accounts for 951, 878, 873, and 868 images for the *b'*, *v*, *w*, and *x* filters, respectively. These images were calibrated in terms of radiance factor (RADF) in line with the Golish et al. (2020) methodology. Their calibration method also removes known sources of noise such as dark current, charge smear, and pixel non-uniformity.

Following the steps of Golish et al. (2021a), we compute the photometric angles by means of Integrated Software for Imagers and Spectrometers 3 (ISIS3; Keszthelyi et al., 2013) and SPICE kernels (Acton et al., 2018) and a shape model provided by the mission. Specifically, we use the same shape model of Bennu, version v28 with a mean ground sample distance (GSD) of 80 cm (Barnouin et al., 2019, 2020). In images where the resolution exceeds that of the shape model GSD, we reduce it to the same value. Otherwise, we could be getting in the same area different reflectance values for the same phase, emission and incidence value, something without any physical meaning. Later, we select pixels following the same criteria followed in Golish et al. 2021a: incidence, or emission angles below 82°, and RADF above 0.001. In addition, we exclude pixels with a phase above 90° due to low signal-to-noise-ratio (SNR). After this last filtering, the number of useful images is reduced to 408, 338, 338 and 338 for the *b'*, *v*, *w*, and *x* bands, respectively.

After data selection, we compute the equigonal albedo dividing the RADF by the disk function. Finally, we average for each image the phase value and the equigonal albedo, and we fit the ROLO phase function to these points with a Levenberg-Marquardt algorithm (Levenberg 1944) (black curve in Figure 2), an iterative procedure widely used to solve non-linear least squares problems. In Table 1, we present the values obtained for the parameters.

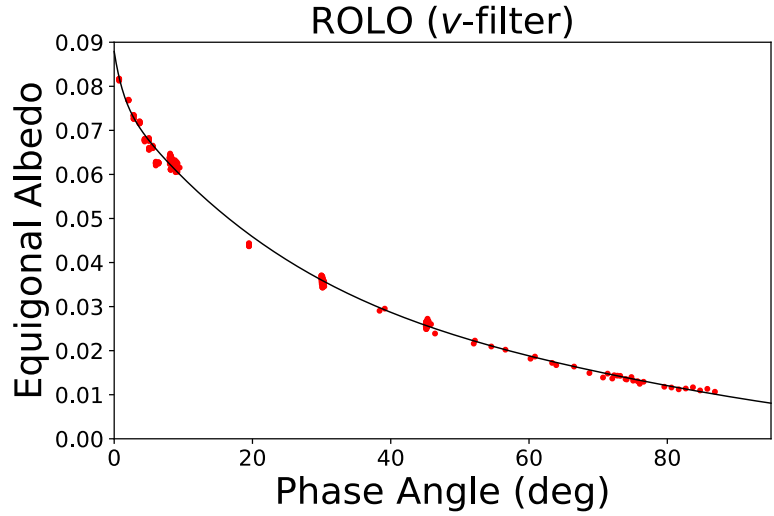

*Figure 2. ROLO phase function (black line) for the v-filter. Red points are the average equigonal albedo of each image. The phase angle ranges from 0° to 90°.*

*Table 1. Values of the photometric parameters for the ROLO phase function used together with the Lommel-Seeliger disk function.*

|  | *b'* | *v* | *w* | *x* |
|---|---|---|---|---|
| $C_0$ | 0.0112 | 0.0107 | 0.0091 | 0.0088 |
| $C_1$ | $7.169 \cdot 10^{-1}$ | $7.336 \cdot 10^{-1}$ | $7.618 \cdot 10^{-1}$ | $7.204 \cdot 10^{-1}$ |
| $A_0$ | $7.633 \cdot 10^{-2}$ | $7.709 \cdot 10^{-2}$ | $7.505 \cdot 10^{-2}$ | $7.275 \cdot 10^{-2}$ |
| $A_1$ | $-1.961 \cdot 10^{-3}$ | $-2.062 \cdot 10^{-3}$ | $-1.951 \cdot 10^{-3}$ | $-1.794 \cdot 10^{-3}$ |
| $A_2$ | $2.576 \cdot 10^{-5}$ | $2.926 \cdot 10^{-5}$ | $2.593 \cdot 10^{-5}$ | $2.097 \cdot 10^{-5}$ |
| $A_3$ | $-1.784 \cdot 10^{-7}$ | $-2.269 \cdot 10^{-7}$ | $-1.828 \cdot 10^{-7}$ | $-1.141 \cdot 10^{-7}$ |
| $A_4$ | $4.598 \cdot 10^{-10}$ | $7.044 \cdot 10^{-10}$ | $4.989 \cdot 10^{-10}$ | $1.609 \cdot 10^{-10}$ |

## 2.2 Artificial Neural Networks

An artificial neural network (ANN) is a mathematical model inspired in the fundamental structure of the brain (Schmidhuber, 2015). ANNs are composed of neurons arranged in layers with arbitrary depth —linked by composite functions— that can learn from data. The most basic ANN consists of one input layer, one hidden layer and an output layer. In an ANN the input data is fed in the forward direction through the network. Each hidden layer accepts the input data, processes it, and passes to the successive layer. In each neuron, the inputs ($x_1, x_2, \ldots x_m$) are converted into an output by means of weights ($w_1, w_2, \ldots, w_m$), a bias ($b$), and a non-linear function called activation function ($f$) as follows:

$$output = f(\sum_{j=0}^{m} w_j x_j + b) \qquad (5)$$

Usually weights and biases are initialized (randomly created) using a normal distribution with fixed standard deviation. However, deep ANN models can have difficulty converging if not properly initialized. For this reason and to avoid this problem, we use the Kaiming method (Kaiming et al. 2015) that gives particularly good results in practice. Kaiming is initialization method for neural networks that consider the non-linearity of activation functions, such as ReLU activations, which avoid reducing or magnifying input signals exponentially.

Activation functions give neural networks their power to model complex non-linear relationships. In this work we use *elu* —to connect each layer— and *sigmoid* —just in the last layer— as the activation functions, which are defined in Eq 6 and 7:

$$f_{ELU}(x) = \begin{pmatrix} x, & if \ x > 0 \\ \alpha(e^x - 1) & if \ x \leq 0 \end{pmatrix} \quad (6)$$

$$f_{SIGMOID}(x) = \frac{1}{1 + e^{-x}} \quad (7)$$

Once the input is propagated through the output layer, the predicted value is compared to the real one (extracted from the dataset, hence the name of *supervised* learning) by a loss function. In this work we use the mean square error (MSE) (Eq. 8), which is widely used for its performance in analogue problems:

$$MSE = \frac{1}{N} \sum_{i=1}^{N} (label_i - prediction_i)^2 \quad (8)$$

The loss is optimized by modifying the weights and biases using Adam stochastic gradient algorithm (Diederik et al. 2015). The gradient is computed by using backpropagation. A sequential iteration of this cycle (epoch) is repeated until a convergence of MSE is reached.

For shorten computing time, we set maximum number of 1100 pixels of each image, which are randomly selected from our data set —it was checked that this data reduction does not affect our final results. To avoid overfitting, the dataset is separated into training and validation data. Validation data are not used for training but are tested for convergence in each epoch. For that, we randomly separate our data into two groups with a 9:1 ratio, i. e., 90% data for training and 10% for validation (Table 2). Figure 3 shows the phase, emission and incidence ranges used in this work.

*Table 2. Number of pixels used in this work for both training and validation. For each pixel we extract the phase, emission, incidence and RADF values, so the total number is four times the values shown in this table.*

|  | **Training** | **Validation** |
|---|---|---|
| ***b'* filter** | 417025 | 46480 |
| ***v* filter** | 352044 | 39255 |
| ***w* filter** | 346944 | 38694 |
| ***x* filter** | 345952 | 38584 |

To evaluate the modeling efficiency, we calculate the MSE value for both the classical and ANN predictions. The percent improvement is given by Eq (9):

$$Improvement\ (\%) = \frac{MSE_{ROLO} - MSE_{ANN}}{MSE_{ROLO}} \cdot 100 \qquad (9)$$

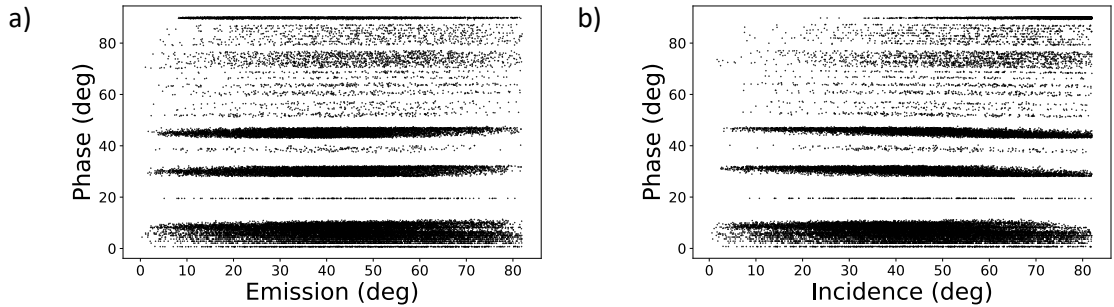

Figure 3. Visualization of the data range used in this work. For this inspection we choose randomly 10% of all v-filter pixels and represent phase against emission (a), and phase against incidence (b).

The input layer of our ANN contains 3 units: phase, emission, and incidence angle values, while the output layer is the measured RADF for these three angles (Figure 4). To implement the ANN, we use the open-source machine learning library PyTorch, which provides accelerated computation using graphical processing units (GPUs), which can do the same calculation ~50 fast than a CPU.

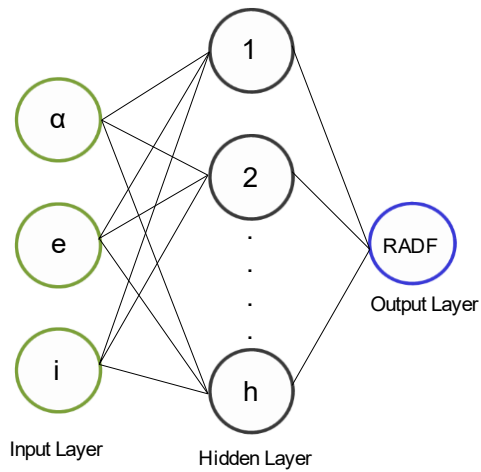

Figure 4. The most basic neural network composed only for one hidden layer.

## 3. Results

The hyperparameters of the ANN (number of hidden layers, neurons in each layer, and the learning rate used for the optimization step for backpropagation) need to be tuned to get optimal performance. To select the best configuration, we made a series of trial-and-error tests where we compute the MSE for a sample of data using a fixed number of 200 epochs. The best set-up will be this one that offers the lowest MSE value but also presents a smooth and constant decrease (Figure 5). For this evaluation we combine a large number of possible configurations, starting from 1 hidden layer and increasing this number, and in parallel modifying the number of neurons —3, 5, 7, 9, 12, 15, 18 or 21— and the Adam learning rate —1, 1E-1, 1E-2, 1E-3, 1E-4, or 1E-5. For more than 5 hidden layers, the computational times increased remarkably while the MSE did not decrease, so it is the maximum number tested here. In Table 3 the best results (lowest MSE values) are shown for each number of hidden layers.

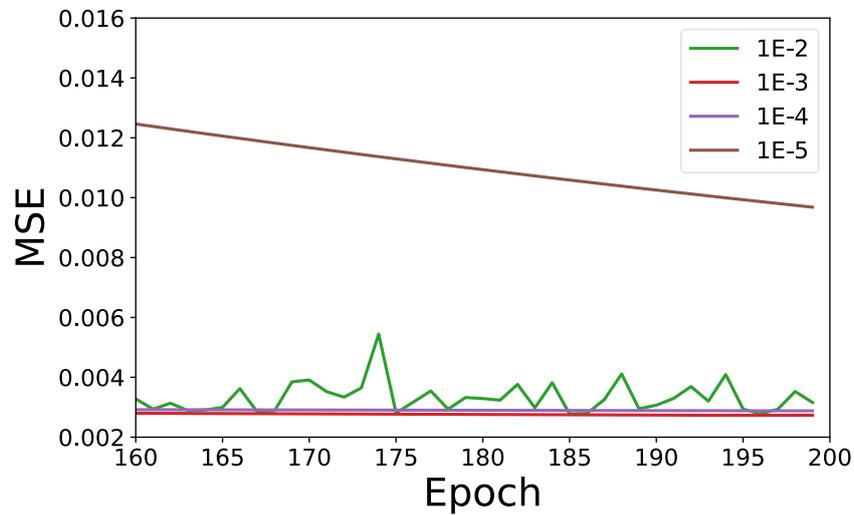

*Figure 5. MSE vs epoch using 3 hidden layers and 15 neurons in each one. For a learning rate larger than 1E-2, the MSE variation is irregular. The lower MSE is reached for a learning rate of 1E-3.*

From this experience we concluded that the best set-up is 3 hidden layers, with 15 neurons in each layer, and a learning rate equal or less than 1E-3.

*Table 3. Test to identify the best ANN configuration running 200 epochs. In this table we show the best case for an increasing number of layers, from 1 up to 5. We also include the computation time (in seconds) but after fixing the learning rate at 1E-4 and the number of neurons at 9 for an appropriate comparison.*

| Hidden Layers | N. of neurons | Learning Rate | MSE | Time (s) |
|:---:|:---:|:---:|:---:|:---:|
| 1 | 7 | 1E-2 | $2.77 \cdot 10^{-3}$ | 14.28 |
| 2 | 9 | 1E-3 | $2.76 \cdot 10^{-3}$ | 20.62 |
| 3 | 15 | 1E-3 | $2.73 \cdot 10^{-3}$ | 26.36 |
| 4 | 5 | 1E-3 | $2.78 \cdot 10^{-3}$ | 32.34 |
| 5 | 7 | 1E-3 | $2.74 \cdot 10^{-3}$ | 37.78 |

After identifying the best ANN, we train our data for each color filter. We stop the training process at 6000 epochs because at this point the MSE remains practically constant, reaching an asymptotic regime. To accelerate the process and reduce computation times taking advantage of the parallel computing GPU platform provided by PyTorch, we separated the total set of training data into batches of 1000 pixels, so in each epoch we calculate, optimize and update weights for each one of them. Finally, we use our trained network to test the modeling efficiency. For that, we use the validation (non-trained) data, and we simulate the RADF using phase, emission, and incidence angles by means of both ROLO model and our trained ANN. We compute the improvement (Eq. 9) through MSE. Our method reaches a more precise modeling for all color filters than ROLO, with an improvement over the classic approach of up to 14.30% for the *x* filter (Table 4). Note that the highest number of training data are available for the *b'* filter, so the efficiency is not correlated with the amount of data used.

*Table 4. MSE value and modeling improvement (in percentage) for each MapCam color filter. These numbers are obtained using validation (non-trained) data.*

|  | $MSE_{ROLO}$ | $MSE_{ANN}$ | Improvement (%) |
| --- | --- | --- | --- |
| *b'* filter | 0.00435 | 0.00373 | 14.15 |
| *v* filter | 0.00453 | 0.00397 | 12.54 |
| *w* filter | 0.00441 | 0.00392 | 11.13 |
| *x* filter | 0.00423 | 0.00362 | 14.30 |

If we represent the predicted RADF as a function of the phase value (fixing emission and incidence angles at 30°), we see how both ROLO and our trained ANN show a similar phase function (Figure 6). However, it follows from our results that more complex relationships than those represented by the ROLO photometric model are presented and they are being captured by our ANN. Also, we can see that the real measurements (overlapped black points in Figure 6) present a high RADF variation for the same photometric angles. Apart from the natural scatter due to noise, it also could be that the heterogeneity of the surface (DellaGiustina et al., 2020), thus suggesting that several photometric models could be needed to describe the scattering properties of different materials over the surface. This is planned as an improvement in our next approach.

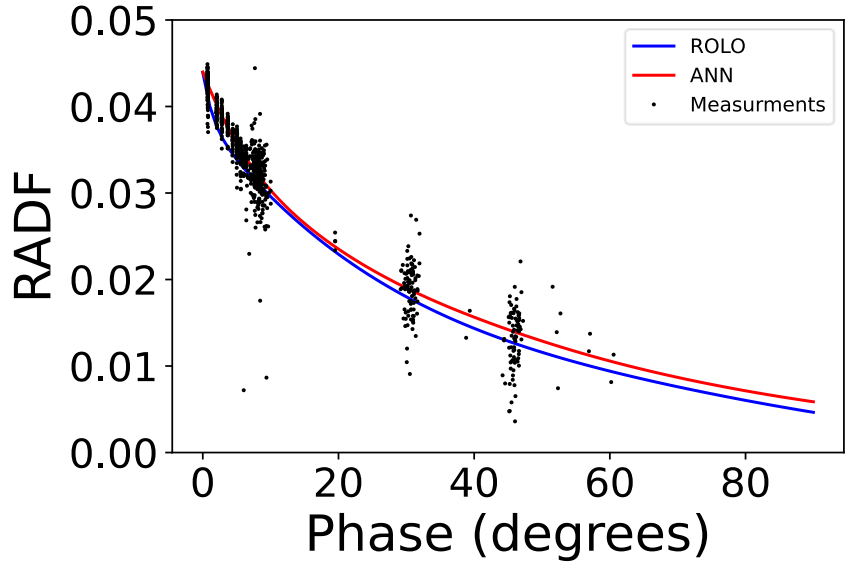

*Figure 6. Simulated RADF values from ROLO model (blue curve) and our trained ANN (red curve) for the phase range 0° - 90°. Overlapping back points are direct measurements used to train the ANN.*

## 4. Conclusions and future work

Photometric modeling is essential when applying corrections before analyzing spectral data from resolved surfaces to remove the effects of varying observation and illumination conditions. In this work we decided to develop a machine learning methodology via an artificial neural network (ANN) to improve the efficiency of photometric modeling. To do this, first we replicated the classical approach, consisting of the search for parameters of previously defined photometric models (Golish et al. 2021a). In parallel, we trained an ANN with the same data used for the classic modeling. Later, to compare both systems, we predicted RADF from phase, emission and incidence angles using non-trained or validation data and compute the MSE in each case. We found that our trained ANN obtains a higher precision model, up to 14.30% for the *x*-filter, in addition to a considerable reduction in time. When we represent both ROLO and ANN predicted phase function curves (Figure 6) along direct measurements, we see that real measurements present a high degree of scatter. This means that we are detecting albedo variations for the same phase, emission, and incidence angles. There are certain natural uncertainties due to errors in shape model, image to shape model misregistration, inaccurate pointing, artifacts or MapCam instrumental noise. However, heterogeneity observed in the surface also suggests the presence of boulders with varying composition, including potentially exogenous material, having originated in other bodies such as (4) Vesta (DellaGiustina et al. 2021). It is likely that these varying materials will have distinct light scattering properties. Thus, a next step will be to consider latitude and longitude values along with phase, emission, and incidence as inputs in our neural network. We expect that regional photometry modelling will further improve our precision.

Moreover, photometric corrections are just application of photometric modelling. As discussed in Section 1, there are many characteristics such as porosity, particle size, or roughness, that can be inferred by studying how light is reflected from a planetary surface. It can be done using Hapke model, which presents a set of parameters with physical meaning. A machine learning approach could be an ideal approach to improve precision, as well as overcoming some of the difficulties of physically motivated models, such as convergence issues or correlations between supposedly independent parameters. Finally, through PDS, we have access to a very wide set of images and shape models of other Solar System bodies acquired by space missions such as Dawn, that visited the asteroid (4) Vesta or the dwarf planet (1) Ceres, or NEAR-Shoemaker, which visited (433) Eros. Therefore, we plan to apply this new methodology to other planetary surfaces to assess the light-scattering behavior of those airless bodies.

## 5. Acknowledgements

We thank the entire OSIRIS-REx Team for making this mission possible. A. Asensio Ramos acknowledges financial support from MICIU through project PGC2018-102108-B-I00 and FEDER funds. J. L. Rizos, J. Licandro, J. de León and M. Popescu acknowledge support from the AYA2015-67772-R (MINECO, Spain). J. de León acknowledges financial support from the Severo Ochoa Program SEV-2015-0548 (MINECO) and the project ProID2017010112 under the Operational Programmes of the European Regional Development Fund and the European Social Fund of the Canary Islands (OP-ERDF-ESF), as well as the Canarian Agency for Research, Innovation and Information Society (ACIISI).

## 6. Data Availability

This research has made use of the USGS Integrated Software for Imagers and Spectrometers (ISIS) (https://isis.astrogeology.usgs.gov/). OCAMS/MapCam images used in this work are available via the Planetary Data System (Rizk et al., 2019). Shape models of Bennu are available via the Small Body Mapping Tool (http://sbmt.jhuapl.edu/). D.N. DellaGiustina and D.R. Golish were supported by NASA under contract NNM10AA11C issued through the New Frontiers Program.